\documentclass[a4paper,11pt]{article}
\pdfoutput=1 
\usepackage[export]{adjustbox}

\usepackage{jcappub}
\usepackage{graphicx}
\usepackage{dcolumn}
\usepackage{amssymb,amsmath,bm}
\usepackage{mathtools}
\usepackage{color}
\usepackage[dvipsnames]{xcolor}
\usepackage{xfrac}
\usepackage{aas_macros}
\usepackage{mathrsfs}
\usepackage{subcaption}
\usepackage{rotating}
\usepackage{chngcntr}
\usepackage[T1]{fontenc} 
\usepackage{natbib}
\usepackage{lipsum}
\usepackage{hyperref}
\usepackage[nameinlink,capitalise]{cleveref}
\newcommand{\nv}{\hat{\bm n}}

\newcommand{\deq}{\coloneqq}

\newcommand{\Ecut}{E_\text{cut}}
\newcommand{\Emax}{E_\text{max}}

\newcommand{\bcut}{b_\text{cut}}
\newcommand{\cN}{\mathcal{N}}
\newcommand{\cS}{\mathcal{S}}

\newcommand{\de}{\mathrm{d}}

\newcommand{\ra}{\rightarrow}

\newcommand{\fp}{f_\text{p}}
\newcommand{\fH}{f_\text{H}}
\newcommand{\fFe}{f_\text{Fe}}

\newcommand{\fmix}{f_\text{mix}}
\newcommand{\ffid}{f_\text{fid}}
\newcommand{\Fmix}{F_\text{mix}}

\newcommand{\Hquad}{\hspace{0.3em}} 

\def\dd{ \text{d} }

\def\+{\dagger}
\def\la{\langle}
\def\ra{\rangle}

\def\n0{{\la n \ra}}

\def\B0{{\la B \ra}}

\title{Honing cross-correlation tools for inference on ultra-high-energy cosmic-ray composition}
\author[a,\,b,\,1]{Konstantinos Tanidis,}
\author[b]{Federico R.\ Urban,}
\author[c,\,d,\,e]{and Stefano Camera}

\affiliation[a]{Department of Physics, University of Oxford,\\Denys Wilkinson Building, Keble Road, Oxford OX1 3RH, UK}
\affiliation[b]{CEICO, Institute of Physics of the Czech Academy of Sciences,\\Na Slovance 1999/2, 182 21 Prague, Czech Republic}
\affiliation[c]{Dipartimento di Fisica, Universit\`a degli Studi di Torino,\\Via P.\ Giuria 1, 10125 Torino, Italy}
\affiliation[d]{INFN -- Istituto Nazionale di Fisica Nucleare, Sezione di Torino,\\Via P.\ Giuria 1, 10125 Torino, Italy}
\affiliation[e]{INAF -- Istituto Nazionale di Astrofisica, Osservatorio Astrofisico di Torino,\\Strada Osservatorio 20, 10025 Pino Torinese, Italy}

\footnotetext[1]{email: konstantinos.tanidis@physics.ox.ac.uk}

\abstract{The chemical composition of the highest-energy cosmic rays, namely the atomic number \(Z\) of rays with energies \(E\gtrsim40~\mathrm{EeV}\), remains to date largely unknown. Some information on the composition can be inferred from the deflections that charged ultra-high-energy cosmic rays experience while they traverse intervening magnetic fields. Indeed, such deflections distort and suppress the original anisotropy in the cosmic ray arrival directions; thus, given a source model, a measure of the anisotropy is also a measurement of the deflections, which in turn informs us on the chemical composition. In this work, we show that, by quantifying ultra-high-energy cosmic ray anisotropies through the angular cross-correlation between cosmic rays and galaxies, we would be able to exclude iron fractions \(f_{\rm Fe}\geq{\cal O}(10\%)\) assuming a fiducial hydrogen map at \(2\,\sigma\) level, and even smaller fractions in the reverse case of hydrogen on an iron map, going well below \(\fH\approx10\%\) when we mask the Galactic Centre up to latitudes of \(40^\circ\). This is an improvement of a factor of a few compared to our previous method, and is mostly ascribable to a new test statistics which is sensitive to each harmonic multipole individually. Our method can be applied to real data as an independent test of the recent claim that current cosmic-ray data can not be reproduced by any existing model of the Galactic magnetic field, as well as an additional handle to compare any realistic, competing, data-driven composition models.}

\begin{document}
\maketitle
\flushbottom

\section{Introduction}\label{sec:intro}

The chemical composition of the highest-energy cosmic rays (UHECRs), namely cosmic rays with energies \(E\gtrsim40\,\mathrm{EeV}\), remains to-date poorly known even after decades of investigations \citep{AlvesBatista:2019tlv,Zhezher:2021qke,Mayotte:2023Nc}. The chemical composition of UHECRs, indeed their atomic number \(Z\), is very hard to measure experimentally on an event-by-event basis, and only aggregate, statistical information is available. Even then, systematic errors on the composition from current data are large and prevent unequivocal determination of UHECR composition as a function of energy, especially at the highest end of the spectrum. Indirect measurements of UHECR composition can be performed by measuring the deflections that charged UHECRs experience while they traverse intervening Galactic and possible extra-Galactic magnetic fields (GMF and xGMF, respectively). In order for these methods to work, one needs to know the distribution of UHECR sources as well as the strength and structure of intervening magnetic fields; neither of which are known well enough at present, unfortunately \citep{Boulanger:2018zrk}. Some useful conclusions can nevertheless be drawn by noticing that the uncertainty in atomic number, which can range from \(Z=1\) for protons (\({^1}\)H) to \(Z=26\) for iron nuclei (\({^{56}}\)Fe), is by far the largest unknown factor that determines the uncertainty in the deflection of UHECRs. This fact was behind recent methods and analyses \citep{Ahlers:2017wpb,dosAnjos:2018ind,Kuznetsov:2020hso,Tanidis:2022jox,Kuznetsov:2023xly}. In particular, in our previous work \citep{Tanidis:2022jox} we have shown that using the harmonic-space cross-correlation power spectrum (often, `angular cross-correlation', for short) between UHECRs and galaxies, if \(Z=56\) we can exclude a 42\% fraction of \({^1}\)H nuclei (or more) at \(2\,\sigma\) for UHECRs above 100~EeV. In this paper we demonstrate how we can hone these tools and improve this result by a factor of~4, excluding as little as 12\% \({^1}\)H using the same synthetic datasets, which we can further restrict to 7\% if only the polar caps at latitudes higher than \(|b|=40^\circ\) are considered.

This paper is structured as follows. In \cref{sec:modelling} we describe the flux model for UHECRs. In \cref{sec:method} we expound our method and detail the test statistics (TS) we employ. In \cref{sec:results} we present our main results, which are then contextualised in \cref{sec:conclusions} together with an outlook for future perspectives. We gather all additional results in \cref{app:some}.

\section{Ingredients}\label{sec:modelling}

Any function \(\Delta_a(\nv)\) of direction \(\nv\) on the celestial sphere can be decomposed into harmonic coefficients as
\begin{align}\label{eq:sph_harm}
	\Delta_{\ell m}^a &\deq \int\dd\nv\,Y^*_{\ell m}(\nv)\,\Delta_a(\nv) \;,
\end{align}
where \(Y_{\ell m}\) are Laplace's spherical harmonics. In this work the functions being correlated are the anisotropy fields of galaxies and UHECRs, for which \(a\in\left\{\text{g},\text{CR}\right\}\), respectively.

\subsection{Galaxies}
Following closely the notation of \citep{Tanidis:2022jox}, we write the anisotropy of the galaxy sample as
\begin{align}\label{eq:g_ani}
    \Delta_\text{g}(\nv) &\deq \int \de\chi\;\phi_\text{g}(\chi)\,\delta_\text{g}(z,\chi\,\nv)\;,
\end{align}
where \(\delta_\text{g}(z,\chi\,\nv)\) is the three-dimensional galaxy overdensity, for which the radial kernel of the galaxy distribution \(\phi_\text{g}(\chi)\) represents the weighted distribution of galaxies as a function of radial comoving distance, \(\chi\), which is related to the cosmological redshift \(z\) through \(\de\chi/\de z=1/H(z)\), with \(H\) the Hubble factor. The galaxy kernel \(\phi_\text{g}(\chi)\) is given by
\begin{align}\label{eq:g_ker}
	\phi_\text{g}(\chi)\deq \left[\int \de\tilde\chi\; \tilde\chi^2\,w(\tilde\chi)\,\bar{n}_{\rm g,c}(\tilde\chi)\right]^{-1}\,\chi^2\,w(\chi)\,\bar{n}_{\rm g,c}(\chi) \;,
\end{align}
where \(\bar{n}_{\rm g,c}(\chi)\) is the comoving, volumetric number density of galaxies in the sample.

The quantity \(w(\chi)\) is an optional distance-dependent weight that can be applied to all the objects in the galaxy catalogue, provided their redshifts are known. Assuming Poisson statistics, it can be shown \citep{Alonso:2020mva} that the optimal weights that maximise the signal for the galaxy-UHECR cross-correlation (XC) are given by
\begin{align}\label{eq:opt_w}
    w(\chi) &= \frac{\alpha(z,\Ecut;\gamma,Z)}{(1+z)\,\chi^2\,\bar{n}_{\rm g,c}(\chi)} \;,
\end{align}
where \(\alpha\) is the attenuation factor, defined as the fraction of cosmic rays injected with \(E\geq\Ecut\) which are detected on Earth with energies greater than \(\Ecut\); the attenuation factor depends upon the atomic number \(Z\) and the injection energy-spectrum, power-law index \(\gamma\) (see below).

We mould our mock galaxy sample on the 2MASS Redshift Survey (2MRS) \citep{Huchra:2011ii}, which is among the most complete full-sky spectroscopic low-redshift surveys, with \(43\,182\) objects \citep[see][]{Ando:2017wff}. Moreover, we limit our mock sample to the minimum distance of \(5\,\mathrm{Mpc}\), as is customary in the UHECR literature, in order for the many-source assumption to be valid \citep{Waxman:1996hp,Koers:2008ba}---were we not to do so, we would also introduce an unphysical cancellation of cosmic variance at low-\(\ell\) in the combined UHECR auto-correlation (AC) and galaxy-UHECR XC analysis \citep[see also][]{Urban:2023tbd}.

\subsection{UHECRs}
Under the assumption that UHECR sources are numerous and steady, we can write the UHECR anisotropy field as
\begin{align}\label{eq:cr_ani}
    \Delta_\text{CR}(\Ecut,\nv;\gamma,Z) &\deq \int\dd\chi\,\phi_\text{CR}(\Ecut,\chi;\gamma,Z)\,\delta_{\rm s}(z,\chi\,\nv) \;,
\end{align}
where the UHECR radial kernel for UHECRs above \(\Ecut\), coming from sources at distance \(\chi\), emitting UHECRs with atomic number \(Z\) and injection spectrum \(\propto E^{-\gamma}\) reads
\begin{align}\label{eq:cr_ker}
    \phi_\text{CR}(\Ecut,z;\gamma,Z) &\deq \left[\int\dd \tilde{z}\,\frac{\alpha(\tilde{z},\Ecut;\gamma,Z)}{H(\tilde{z})\,(1+\tilde{z})}\right]^{-1} \frac{\alpha(z,\Ecut;\gamma,Z)}{(1+z)} \;.
\end{align}
Here, \(\delta_{\rm s}(z,\chi\,\nv)\) is the UHECR source density contrast \citep[for derivation and further references, see][]{Urban:2020szk}), and the attenuation function \(\alpha(z,\Ecut;\gamma,Z)\) accounts for the probability that a UHECR detected with energy above \(\Ecut\) had originated from a source located at redshift \(z\). The attenuation function, taking into account all energy losses, is calculated with the propagation code \emph{SimProp} v2r4 \citep{Aloisio:2017iyh}. For nuclei injected with \(E>A\,\Ecut\) where \(A\) is the mass number, we take into account secondary, ``child'' lighter nuclei (which we count as protons) \citep[see][]{Tanidis:2022jox}. 

Assuming that UHECRs are of one given species \(Z_1\), for instance \({^1}\)H with \(Z_1=1\), we seek to determine what is the fraction \(\fmix\) of a second species \(Z_2\), e.g.\ \({^{56}}\)Fe with \(Z_2=26\), that can be constrained through the XC for UHECR datasets of size comparable to what is currently available. Hence, the radial kernel for the \(Z_1\)-\(Z_2\) admixture becomes
\begin{align}\label{eq:cr_mix}
    \phi_\text{CR}(\Ecut,z;\gamma,\fmix) &\propto \frac{(1-\fmix)\,\alpha(z,\Ecut;\gamma,Z_1)+\fmix\,\alpha(z,\Ecut;\gamma,Z_2)}{(1+z)} \;.
\end{align}
In this paper, we specifically want to know how firmly current data can rule out \({^{56}}\)Fe if UHECRs are \({^1}\)H---hence \(\fmix=\fFe\) (or \({^1}\)H if UHECRs are \({^{56}}\)Fe). Notice that this test is not symmetric under the exchange \({^1}\)H\(\leftrightarrow\)\({^{56}}\)Fe, and we will show both cases below.

Owing to the fact that the UHECRs flux is very low---a typical dataset for \(E\gtrsim40\,\mathrm{EeV}\) contains \({\cal O}(10^3)\) events---we can presume that on average each UHECR comes from a different source, and that all UHECR sources are in the galaxy catalogue, such that \(\delta_\text{g}(z,\chi\,\nv) = \delta_{\rm s}(z,\chi\,\nv)\). This is accurate provided that there are many more sources than events \citep{Koers:2008ba}, which is the case in this work.

\subsection{Correlators}
The harmonic-space (or, angular) power spectrum \(\cS^{ab}_\ell\) is the two-point function of the harmonic expansion coefficients of the anisotropy fields, i.e.\
\begin{align}\label{eq:aps}
	\langle \Delta^a_{\ell m}\,\Delta^{b\ast}_{\ell'm'}\rangle &\deq \delta_{\ell\ell'}\,\delta_{mm'}\,\cS^{ab}_\ell \;,
\end{align}
where the angle brackets stand for the ensemble average. The power spectrum \(\cS_\ell^{ab}\) is most conveniently expressed in terms of the three-dimensional Fourier-space power spectrum \(P_{ab}(z,k)\) as
\begin{equation}\label{eq:cl_limber}
	\cS^{ab}_\ell=\int \frac{\de\chi}{\chi^2}\;\phi_a(\chi)\,\phi_b(\chi)\,P_{ab}\left[z(\chi),k=\frac{\ell+1/2}{\chi}\right] \;,
\end{equation}
where the theoretical \(P_{ab}(z,k)\) is modelled according to the halo-model prescription, adapted to the specifics of the 2MRS catalogue \citep{Peacock:2000qk,Ando:2017wff}.

The observed power spectrum is the combination of signal and noise
\begin{equation}\label{eq:cell}
    C^{ab}_\ell \deq \cS^{ab}_\ell+\cN^{ab}_\ell \;.
\end{equation}
Since the angular positions of the UHECRs and the galaxies are discrete point processes,
\begin{equation}\label{eq:noise}
	\cN^{ab}_\ell=\frac{\bar{N}_{\Omega,a\wedge b}}{\bar{N}_{\Omega,a}\,\bar{N}_{\Omega,b}} \;,
\end{equation}
where \(\bar{N}_{\Omega,a}\) is the angular number density of points in sample \(a\), and \(\bar{N}_{\Omega,a\wedge b}\) is the angular number density of points shared in common \citep{Urban:2020szk}. Notice that, since we assume that each UHECR comes from a galaxy in our sample, we can neglect the noise term in the cross-correlation power spectrum.

\subsection{Magnetic fields}
To account for the deflections caused by the intervening GMF, we adopt a data-driven prescription \citep{Pshirkov:2013wka}. In harmonic space, it amounts to applying a scale-dependent beam
\begin{equation}\label{eq:beam_har}
    {\cal B}_\ell \simeq \exp\left[-\frac{\ell\,(\ell+1)\,\sigma^2}{2}\right] \;,
\end{equation}
to the UHECR harmonic coefficients \(\Delta^\text{CR}_{\ell m}\). The width of the beam, \(\sigma\), is given by
\begin{align}\label{eq:smear}
    \sigma \deq \frac{1}{\sqrt2}\left(\frac{40\,\mathrm{EeV}}{E/Z}\right) \frac{1^\circ}{\sin^2 b + 0.15} \;,
\end{align}
where \(b\) is galactic latitude \citep[see][but notice the factor \(\sqrt2\) difference with their normalisation]{Pshirkov:2013wka,diMatteo:2017dtg}.\footnote{Notice that this model accounts for the small-scale GMF only, and ignores the large-scale component of the GMF, which can be comparable or even larger. Nonetheless, we believe our results will hold in a more realistic model of the GMF for three reasons. (1) The model \autoref{eq:smear} is an upper limit on the turbulent deflections from small-scale fields. (2) We always maximise the deflections within a certain area, so we are overestimating the effects of the small-scale GMF. (3) The large-scale GMF acts coherently on the deflections and does not suppresses the global anisotropy in direct proportion to its strength (unlike the small-scale GMF), so its effects on the angular, harmonic correlators are less significant than those of the small-scale GMF.} In practice, in order to keep with a mostly analytic treatment of the deflections, we conservatively smear uniformly across the sky with the largest \(\sigma\) in a given region of the sky. For instance, if we keep a full sky we set \(b=0\) in \cref{eq:smear}, whereas if we mask the Galactic plane out to some \(\bcut\) we set \(b=\bcut\). For simplicity we neglect any contribution from hypothetical xGMFs in what follows.

Lastly, in order to account for the energy-dependence of the magnetic deflections, we bin the UHECR flux in five logarithmic bins of width~0.1, according to a model differential energy spectrum given by
\begin{equation}\label{eq:en_spectrum}
    J(E)=
    \begin{cases}\displaystyle{J_0\, \left(\frac{E}{\mathrm{EeV}}\right)^{-\gamma_1}} & E \leq E_1 \\
    \displaystyle{J_0\, \left(\frac{E_1}{\mathrm{EeV}}\right)^{-\gamma_1}\,\left(\frac{E}{E_1}\right)^{-\gamma_2}} & E > E_1
    \end{cases}
    \;,
\end{equation}
where \(\gamma_1 = 3\), \(\gamma_2 = 5\), \(E_1 = 10^{19.75}\,\mathrm{eV}\), and \(J_0 = 4.28\times10^6/\mathrm{EeV}\) are our reference values. The total number of events above \(\Ecut\) is approximately 1000, 200, 30 events for \(\Ecut=10^{19.6}\,\mathrm{eV}\approx40\,\mathrm{EeV}\), \(\Ecut=10^{19.8}\,\mathrm{eV}\approx63\,\mathrm{EeV}\), \(\Ecut=10^{20}\,\mathrm{eV}=100\,\mathrm{EeV}\), respectively. These are representative of the number of events we can expect for a full-sky observatory with statistics comparable to existing observatories \citep{Ivanov:2021mkn,PierreAuger:2021ibw}.

\section{Method}\label{sec:method}

In our first work \citep{Tanidis:2022jox}, we developed a test to determine how confidently a full-sky UHECR experiment, which observes a number of events comparable to existing data, can exclude a fraction \(\fFe\) of iron in a pure proton flux (or, vice versa, a fraction \(\fH\) of protons in a pure iron sky) using the UHECR AC and UHECR-galaxies XC power spectra. The method follows closely our previous work, however with some differences that are responsible for our improved results. The method consists of the following steps:
\begin{enumerate}
    \item We choose our fiducial model \(\gamma=2.3\) and \(E_\text{cut}=100\,\mathrm{EeV}\)---we later will repeat the analysis for \(E_\text{cut}=63\,\mathrm{EeV}\) and \(E_\text{cut}=40\,\mathrm{EeV}\). We compute the synthetic data vector at a given $\fmix=\ffid$ (with \(\ffid\in\{0,1\}\) for \({^1}\)H and \({^{56}}\)Fe fiducials, respectively), namely $d^{ab}_\ell=\mathcal{S}^{ab}_{\ell,\ffid}$, and its corresponding standard deviation $\sigma^{ab}_\ell$ which accounts for the observational error due to shot-noise and cosmic variance that is\footnote{Assuming that \(\Delta^a_{\ell m}\) is Gaussian, the covariance matrix of the harmonic-space power spectrum (which is a correlator of four-point statistics) can be written using Wick's theorem. This is expressed as the sum of all the available two-point correlator combinations.}
    \begin{equation}
    \sigma_\text{AC}\equiv\sigma^{\text{CR\,CR}} = \sqrt{\frac{2}{2\ell+1}}C^{\text{CR\,CR}}_\ell\;, \quad \sigma_\text{XC}\equiv\sigma^{\text{g\,CR}} = \sqrt{\frac{\left({\cS^{\text{g\,CR}}_\ell}\right)^2+C^{\text{g\,g}}_\ell\, C^{\text{CR\,CR}}_\ell}{2\ell+1}}\;.
    \label{eq:StdforHistXC}
    \end{equation}
    In addition, we compute the theory vector $t^{ab}_\ell(\Fmix)=\mathcal{S}^{ab}_{\ell,\Fmix}$ for a single parameter $\Fmix$ running in a flat prior in $[0, 1]$ with a step of 0.001. We can define the $\chi^2$ TS
    \begin{equation}
    \chi^2(\Fmix)\deq\sum_\ell\frac{[t_\ell(\Fmix)-d_\ell]^2}{\sigma_\ell^2} \;,
        \label{eq:chi2}
    \end{equation}
    that is expected to be minimised at $\Fmix=\ffid$, which is the value that reproduces the synthetic data (we suppress the \({a,b}\) superscripts henceforth in this section to reduce clutter). The $\chi^2$ values correspond to confidence levels (hereafter CLs) for 1 degree of freedom (that is, for the single parameter $\Fmix$). Then, given the $\Fmix$ for which we are \(q\) CLs away from the synthetic data at $\ffid$, we compute the harmonic-space spectra corresponding to those values as follows:
    \begin{align}
    \text{lower}_{\ell,\ffid, q}&\deq\cS_{\ell,\ffid}-q\,\left|\cS_{\ell,\ffid}-\cS^{\ffid}_{\ell,\Fmix \Hquad \text{at} \Hquad q=1}\right|
    \label{eq:lowerCL}\;,\\
    \text{upper}_{\ell,\ffid, q}&\deq\cS_{\ell,\ffid}+q\,\left|\cS_{\ell,\ffid}-\cS^{,\ffid}_{\ell,\Fmix \Hquad \text{at} \Hquad q=1}\right|\;.
    \label{eq:upperCL}
    \end{align}
    The $q=1,..,5$ define CL bands above and below the spectrum at $\fmix=\ffid$. In \cref{fig:chi2} we present the $\chi^2$ values of the $\Fmix$ model parameter for three scenarios of synthetic data at $\ffid=0$ (in blue colour), $\ffid=0.5$ (in orange colour), and $\ffid=1.0$ (in green colour). Notice that, as anticipated, the \(\ffid=1\) and \(\ffid=0\) are not symmetric.
    \begin{figure}
        \centering
        \includegraphics[width=0.75\columnwidth]{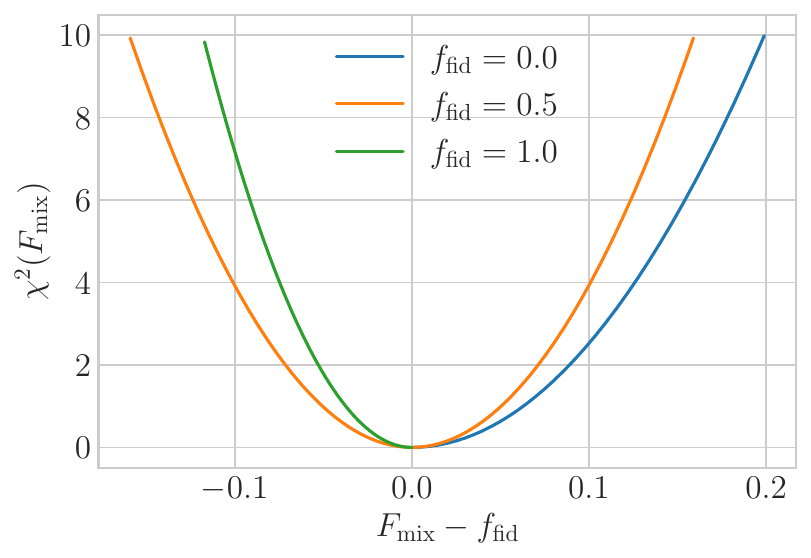}
        \caption{Distribution of $\chi^2$ values as a function of the model parameter $\Fmix$ for synthetic data at $\ffid=0$ (blue), $\ffid=0.5$ (orange) and $\ffid=1.0$ (green). The $\chi^2$ best-fit corresponds to the data value of $\fmix$ as expected.}
        \label{fig:chi2}
    \end{figure}
    \item We compute $N_\text{rea}=1,..,1000$ realisations of the spectra $\mathcal{G}_{\ell, N_\text{rea},\ffid}$ derived from a Gaussian distribution with mean at $\cS_{\ell,\ffid}$ and standard deviation of $|\cS_{\ell,\ffid}-\cS^{\ffid}_{\ell,\Fmix\;\text{at}\;q=1}|$. For each realisation we compute the fraction of points that are between
    \begin{equation}
        \text{lower}_{\ell,\ffid, q} \leq \mathcal{G}_{\ell, N_\text{rea},\ffid} \leq \text{upper}_{\ell,\ffid, q}\;,
        \label{eq:inequalities}
    \end{equation}
    for all \(q\) and \(\ell\) values.
    \item Next, for a given value of the test model parameter, in our case \(\fmix\) (more specifically \(\fFe\), for which we use a step 0.025 in [0, 1]), we compute the \cref{eq:chi2}, obtain the spectra at $\text{lower}_{\ell,\ffid, q}$ and $\text{upper}_{\ell,\ffid, q}$, and repeat the previous steps to obtain the fraction of realisations of the fiducial spectra $\mathcal{G}_{\ell, N_\text{rea},\ffid}$ lie outside the band defined by \cref{eq:inequalities} at the desired \(q\) CL. This represents the confidence with which each realisation of a hypothetical experiment (with its number of detected events and angular resolution) is able to reject the test model.
    \item Overall, this provides the percentage \(n(\fmix;q)\) of experiments that are able to exclude \(\fmix\) at at least \(q\) CL, or---the information can also be read across \(\fmix\) as well as across \(q\) ---exclude \(\fmix\) or more (when the fiducial is \(\ffid=0\), otherwise \(\fmix\) or less if the fiducial is \(\ffid=1\)) at \(q\) CL.
\end{enumerate}

The main difference with respect to the method outlined in \citep{Tanidis:2022jox} is that here we choose to make use of the full harmonic-space power spectrum instead of compressing it to just on value, namely the total power across all multipoles. Having designed a TS that is separately sensitive to each multipole is the main reason behind its improved sensitivity, as we shall see in \cref{sec:results}.

Another difference is that in \citep{Tanidis:2022jox} we generated \(250{,}000\)~realisations of the power spectra, whereas here we find that 1000 realisations are enough to make our predictions stable against fluctuations. This is due to the fact that since each multipole is separately tested against the fiducial, in effect we are generating several hundreds realisations of observables that we are using in the TS (the highest multipoles have no constraining power because of the magnetic beam).


\section{Results}\label{sec:results}

\begin{figure}[thb]
    \centering
    \includegraphics[width=\columnwidth]{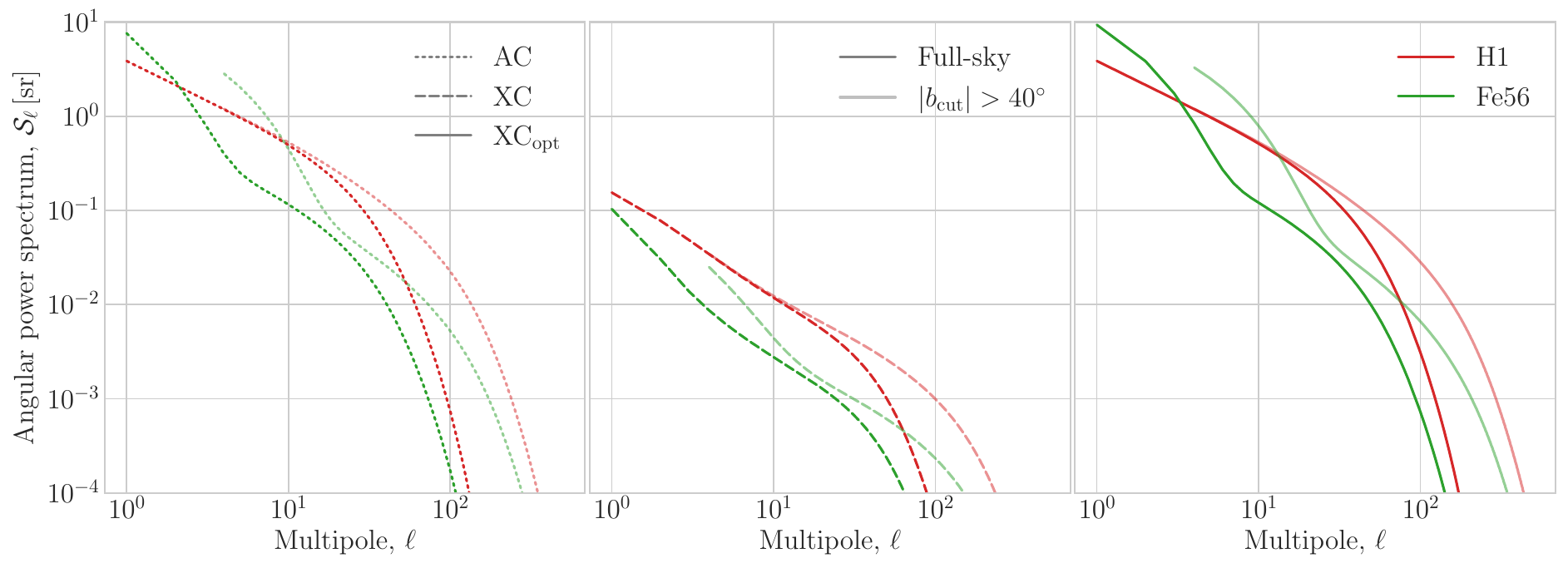}
    \caption{harmonic-space power spectra for \({^1}\)H (red) and \({^{56}}\)Fe (green) at \(\Ecut=100\,\mathrm{EeV}\); the AC (dotted) is on the left panel, the XC (dashed) is on the middle panel and the XC$_\text{opt}$ (solid) is on the right panel. Full-sky results are in dark tones, whereas the results from masking the Galactic Centre at $|\bcut|=40^\circ$ are in lighter tones.}
    \label{fig:spectra}
\end{figure}

In order to gain some intuition as to the UHECR anisotropy as seen by different primaries, in \cref{fig:spectra}, left panel, we show the predicted harmonic-space power spectra for \({^1}\)H (red) and \({^{56}}\)Fe (green) at \(\Ecut=100\,\mathrm{EeV}\) for a full-sky analysis or in the case in which we mask the Galactic Centre at $|\bcut|=40^\circ$.  The \({^{56}}\)Fe curves exhibit a higher anisotropy at large scales (small \(\ell\)), which drops off rapidly at intermediate scales. This is expected because \({^{56}}\)Fe has a slightly shorter radial kernel than \({^1}\)H and, before the magnetic deflections set in at intermediate \(\ell\), it carries a stronger anisotropic imprint than \({^1}\)H. Then, because of the stronger impact of the magnetic deflections, the \({^{56}}\)Fe spectra drop off rapidly. The hardening of the \({^{56}}\)Fe spectra at large \(\ell\) is a result of the child protons developed from an initial pure \({^{56}}\)Fe injection.\footnote{Had we chosen to place a high-energy cutoff in the injection spectrum at \(\Emax=A\Ecut\) for \({^{56}}\)Fe nuclei, there would be in effect no child protons above \(\Ecut\) on Earth. In this case the difference between \({^1}\)H and \({^{56}}\)Fe would be starker, and telling them apart easier, since the \({^{56}}\)Fe spectrum would keep decaying rapidly at high \(\ell\). We show the results for such case in \cref{app:some}.} Also, notice that the difference between the AC and the XC\(_\text{opt}\) is entirely caused by the fact that the magnetic and resolution beams are different for the UHECR map and the galaxy overdensity map; indeed, the unbeamed spectra are identical by construction of the optimal weights. Lastly, note that the spectra for $|\bcut|=40^\circ$ are starting at higher $\ell$ compared to those for the full-sky. This is due to the fact that the minimum multipole $\ell$ we can probe is defined by $\ell_\text{min}(b_\text{cut})\sim\pi/[2f_\text{sky}(b_\text{cut})]$ and the sky fraction is given by $f_\text{sky}(b_\text{cut})=1-\sin(b_\text{cut})$. 

\begin{figure}[thb]
    \centering
    \includegraphics[width=\textwidth]{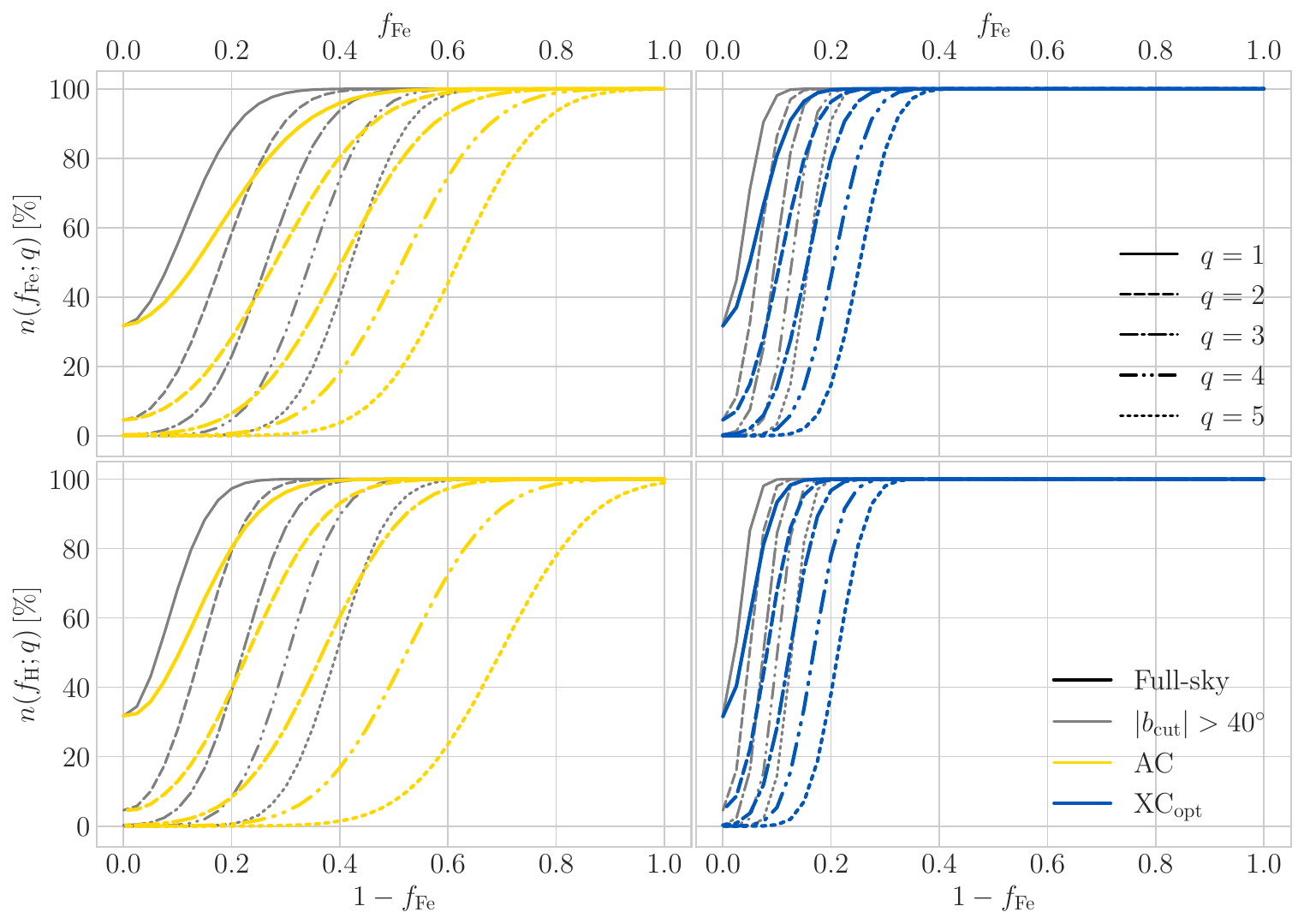}
    \caption{(Top) Percentage of experiments \(n(f_{\rm Fe};q)\) that will be able to exclude \(f_{\rm Fe}\) or more at \(q\) CL for different choices of \(q\): from \(q=1\) (solid) to \(q=5\) (dotted), for our benchmark model with \(\Ecut=100\,\mathrm{EeV}\) and \(\gamma=2.3\). The AC angular power spectra are shown in yellow on the left panel, and the XC\(_\text{opt}\) optimal angular power spectra are shown in blue on the right panel. Darker tones show results for the full-sky, lighter tones show results with the Galactic Centre masked off. (Bottom) Same but exchanging fiducial and test model: in this case we show \(n(f_{\rm H};q)\) and test a fraction \(f_{H}=1-f_{\rm Fe}\) of \(Z=1\) injection against a \(Z=26\) fiducial.}
    \label{fig:Ecuts_fFe}
\end{figure}

In order to assess the strength of our method in discriminating different chemical compositions at high energies, in \cref{fig:Ecuts_fFe}, top panels, we show the percentage of hypothetical experiments \(n(f_{\rm Fe};q)\) that will be able to exclude \(f_{\rm Fe}\) or more at \(q\) CL for our benchmark model with \(\Ecut=100\,\mathrm{EeV}\) and \(\gamma=2.3\). In the bottom panels we show results when we exchange fiducial and test model, namely \(\fH\deq1-\fFe\). Our results show that with the AC we are able to exclude \(\fFe\approx0.40\) (\(\fFe\approx0.27\)) in \(n=80\%\) (\(n=50\%\)) of the tests at \(q=2\)\,CL. These values significantly decrease in the case of the XC\(_\text{opt}\) for which we find \(\fFe\approx0.15\) (\(\fFe\approx0.11\)) in \(n=80\%\) (\(n=50\%\)) of the tests. If we adopt \({^{56}}\)Fe as our fiducial model we find \(\fH\approx0.33\) (\(\fH\approx0.23\)) for the AC and \(\fH\approx0.12\) (\(\fH\approx0.08\)) for the XC\(_\text{opt}\). We do not find a strong dependence on the energy cut \(\Ecut\), see \cref{app:some}. While the qualitative behaviour of these tests reflects that of our previous study \citep{Tanidis:2022jox}, quantitatively the new method described here brings about a significant improvement. For example, using the total angular power summed over \(\ell\in[1,1000]\) in our previous work we found that the XC\(_\text{opt}\) could only exclude \(\fFe\approx0.55\) (\(\fFe\approx0.39\)) in \(n=80\%\) (\(n=50\%\)) of the tests (and similarly for the other cases)---in comparing the two works by eye, notice that here the x-axes run always from \(\fmix=0\) to \(\fmix=0.5\), whereas in \citep{Tanidis:2022jox} they run up to~\(\fmix=1\).

Next, in order to assess how beneficial it can be to adopt a more refined version of the magnetic smearing that takes into account its latitude-dependence (namely, that at higher latitudes the magnetic field has a smaller impact), in lighter tones in \cref{fig:Ecuts_fFe} we re-do the analysis after masking the Galactic Centre within \(|\bcut|=40^\circ\). In this case the smearing angle we adopt is much smaller (see Eq.\ \ref{eq:smear}), and for simplicity (and conservatively) we discard any anisotropic power arising from the masked region. Despite the fact that there is a loss of UHECR events due to the mask, the reduction in magnetic smearing more than makes up for it, improving the XC\(_\text{opt}\) to \(\fFe\approx0.09\) (\(\fFe\approx0.07\)) in \(n=80\%\) (\(n=50\%\)) of the tests for an \({^1}\)H fiducial and \(\fH\approx0.07\) (\(\fH\approx0.05\)) for an \({^{56}}\)Fe fiducial; similar improvements are seen in the AC case. In \cref{app:some} we show results with \(|\bcut|=20^\circ\) and \(|\bcut|=60^\circ\), where it is visible that the best case scenario, but only marginally so, appears to be between 40\,deg and 60\,deg, in rough agreement with the analysis of \citep{Urban:2023tbd}. 

\section{Discussion and outlook}\label{sec:conclusions}

In this paper we have presented an improved method that aims at statistically distinguishing UHECR atomic numbers \(Z\), by quantifying their expected deflections in the GMF. In particular, in this work we build upon our earlier results in \citep{Tanidis:2022jox} where we employed the harmonic, angular AC and XC (specifically, the optimised XC\(_\text{opt}\)) power spectra to quantify UHECR anisotropies, and the impact of the GMF on them. With our improved method, by looking at the XC\(_\text{opt}\) we are able to exclude \({^{56}}\)Fe fractions \(f_{\rm Fe}\leq{\cal O}(10\%)\) on a fiducial \({^1}\)H map at \(2\,\sigma\) level in most of the tests, and even less in the reverse case of \({^1}\)H on a \({^{56}}\)Fe map, in some cases going below \(\fH\approx10\%\) when we mask the Galactic Centre up to \(|\bcut|=40^\circ\). The AC is much less performing, as it is less sensitive to the effects of the GMF (it picks up anisotropies at larger scales than the XC, where the GMF is not as relevant).

These are improvements of a factor of a few compared to our previous method. The main driver behind this improvement is the use of each individual multipole \(\ell\) separately in building the TS, as opposed to aggregating them in the total harmonic, angular power as done in \citep{Tanidis:2022jox}. In this way, and the more so for the XC than the AC, we are able to separately capture the effects of the magnetic deflections at different scales; since the XC is sensitive to a wider range of scales than the AC, its constraining power is also stronger.

In this work we have chosen to work with \({^1}\)H and \({^{56}}\)Fe primarily because, at the highest energies where angular anisotropies are expected to be the most prominent, we do not expect a large contribution of other nuclei, simply because the propagation horizon of intermediate nuclei is much shorter (thereby reducing the available number of sources). Moreover, this work serves to illustrate our method, which then can be rapidly applied to compare more realistic, data-driven chemical compositions. We plan to address this point in future work.

There is some tentative evidence that, using an approach similar to our own, UHECR data is not reproducible with models of the GMF alone \citep{Kuznetsov:2023xly} (see also \citep{Allard:2021ioh,Allard:2023uuk}) and requires additional smearing from, i.e., extra-Galactic magnetic fields. Our method enables a valuable complementary test for this result, as it is designed to detect anisotropies in harmonic-space, thereby further sharpening our understanding of extreme-energy cosmic-ray sources as well as Galactic and astrophysical magnetism.

\acknowledgments
KT is supported by the STFC grant ST/W000903/1 and by the European Structural and Investment Fund. For most of the development of this project KT and FU were supported by the European Regional Development Fund (ESIF/ERDF) and the Czech Ministry of Education, Youth and Sports (MEYS) through Project CoGraDS~-~\verb|CZ.02.1.01/0.0/0.0/15_003/0000437|. SC acknowledges the `Departments of Excellence 2018-2022' Grant awarded by the Italian Ministry of University and Research (\textsc{mur}, L.\ 232/2016).

\appendix
\counterwithin{figure}{section}

\section{Additional results}\label{app:some}

In this appendix we present additional results that show how the sensitivity of our method to the choice of energy cuts, different Galaxy masks and injected maximum energy. Finally, we show how our results can be improved ever so slightly by employing the combination of all data, namely the AC together with the (optimised) XC\(_\text{opt}\) in one overall data vector.

\begin{figure}
    \centering
    \includegraphics[width=\textwidth]{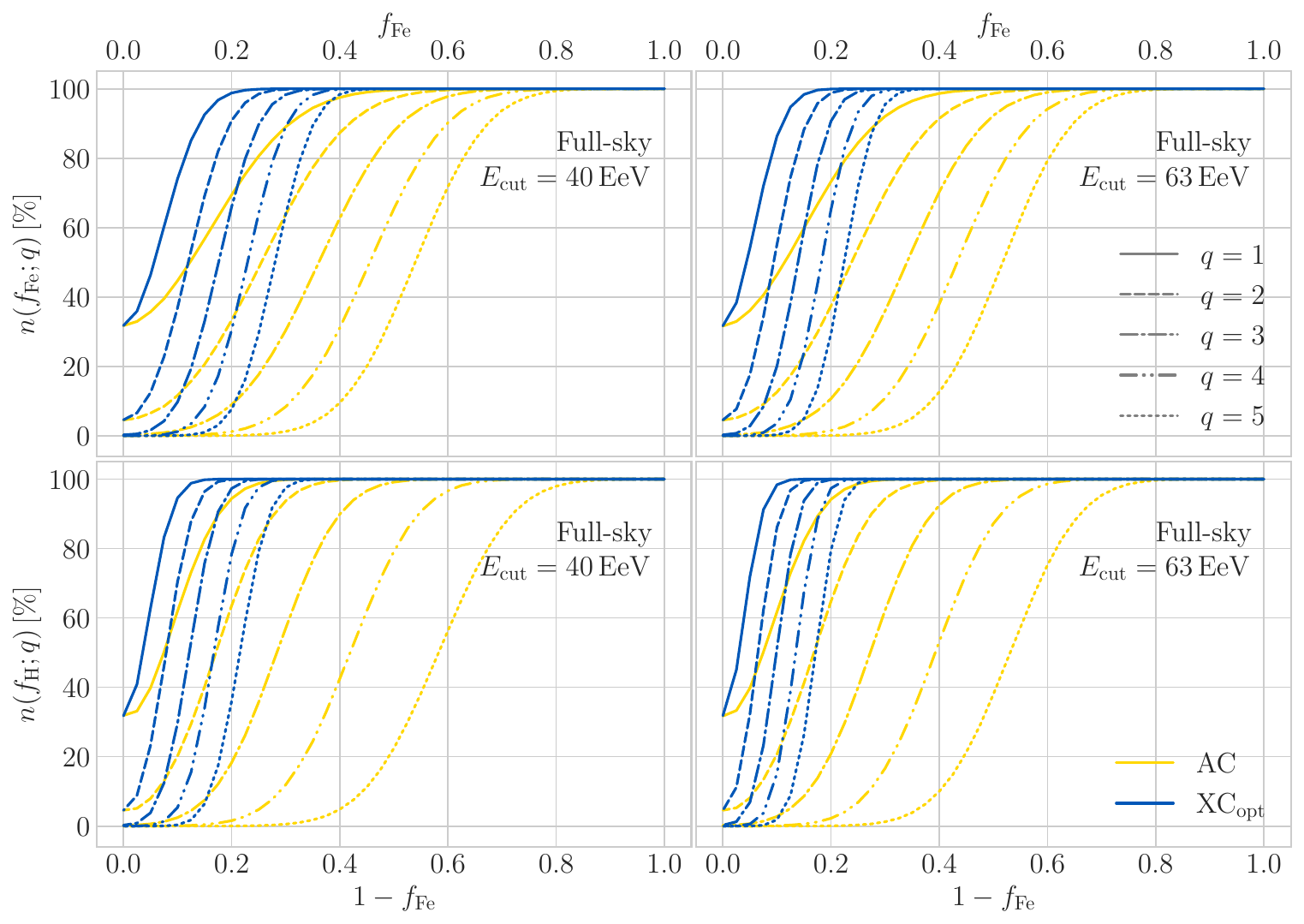}
    \caption{Same as Figure \ref{fig:Ecuts_fFe} but with energy cuts at \(\Ecut=40\,\mathrm{EeV}\) (left) and \(\Ecut=63\,\mathrm{EeV}\) (right), and overlaying AC and XC\(_\text{opt}\) in the same panel.}
    \label{fig:Ecuts_fFe_Ecuts}
\end{figure}

Firstly, in order to assess the dependence on the energy cut \(\Ecut\), in \cref{fig:Ecuts_fFe_Ecuts} we show the changes in constraining power when we adopt \(\Ecut=40\,\mathrm{EeV}\) (left) and \(\Ecut=63\,\mathrm{EeV}\) (right). We do not observe any major quantitative differences, but it is seen that lower energies are less constraining than higher energies. This is in line with our previous results and justifies our choice of \(\Ecut=100\,\mathrm{EeV}\) for our benchmark scenario. The bottom line is that, the better we are able to detect UHECR anisotropies over the noise, the stronger is the constraining power of our method. In this case the loss of UHECR events due to the higher energy cut is more than compensated by the much smaller deflections due to the GMF.

\begin{figure}
    \centering
    \includegraphics[width=\textwidth]{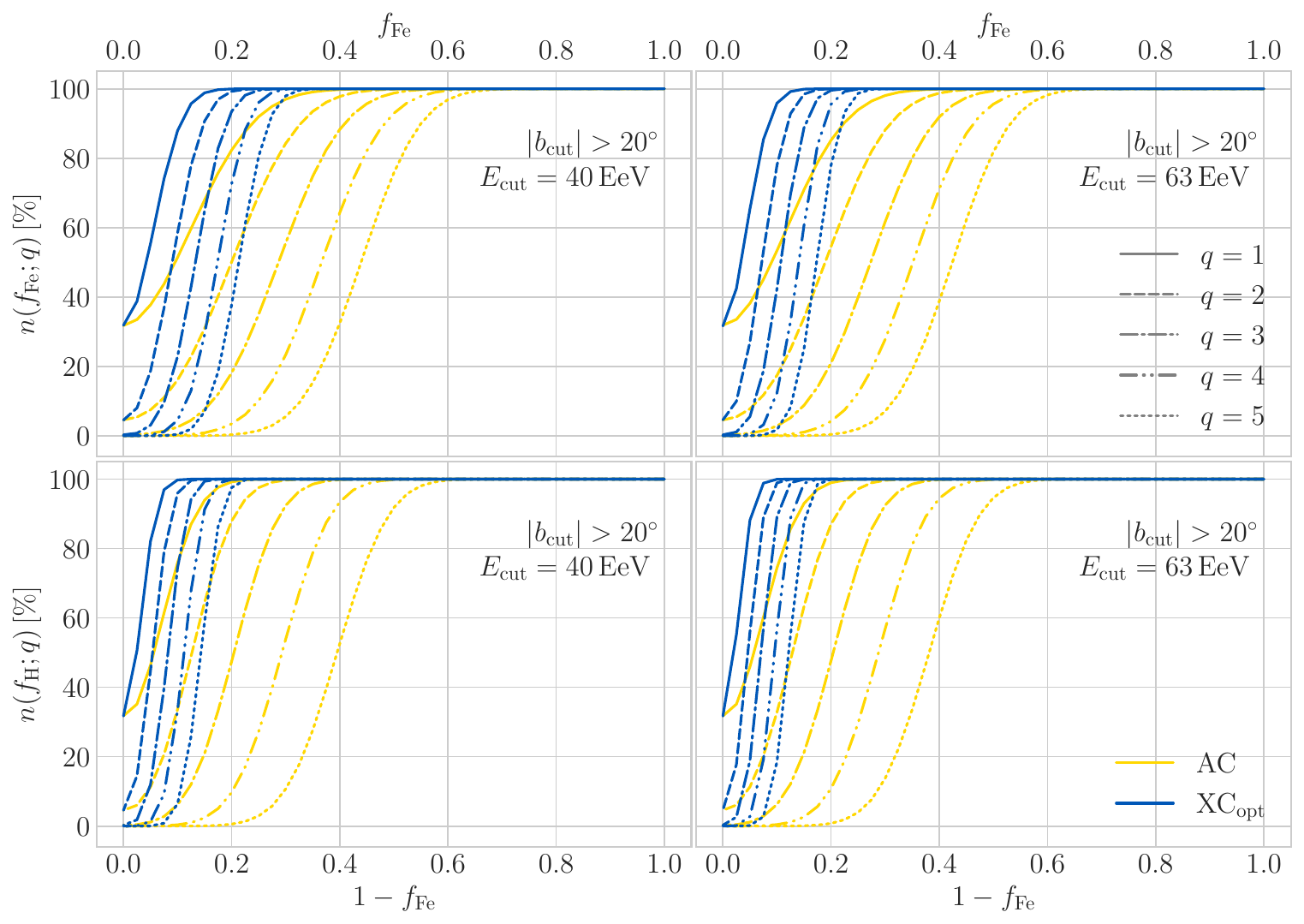}
    \caption{Same as Figure \ref{fig:Ecuts_fFe_Ecuts} but with masking at $|\bcut|=20^\circ$.}
    \label{fig:Ecuts_fFe_Ecuts_mask20}
\end{figure}

\begin{figure}
    \centering
    \includegraphics[width=\textwidth]{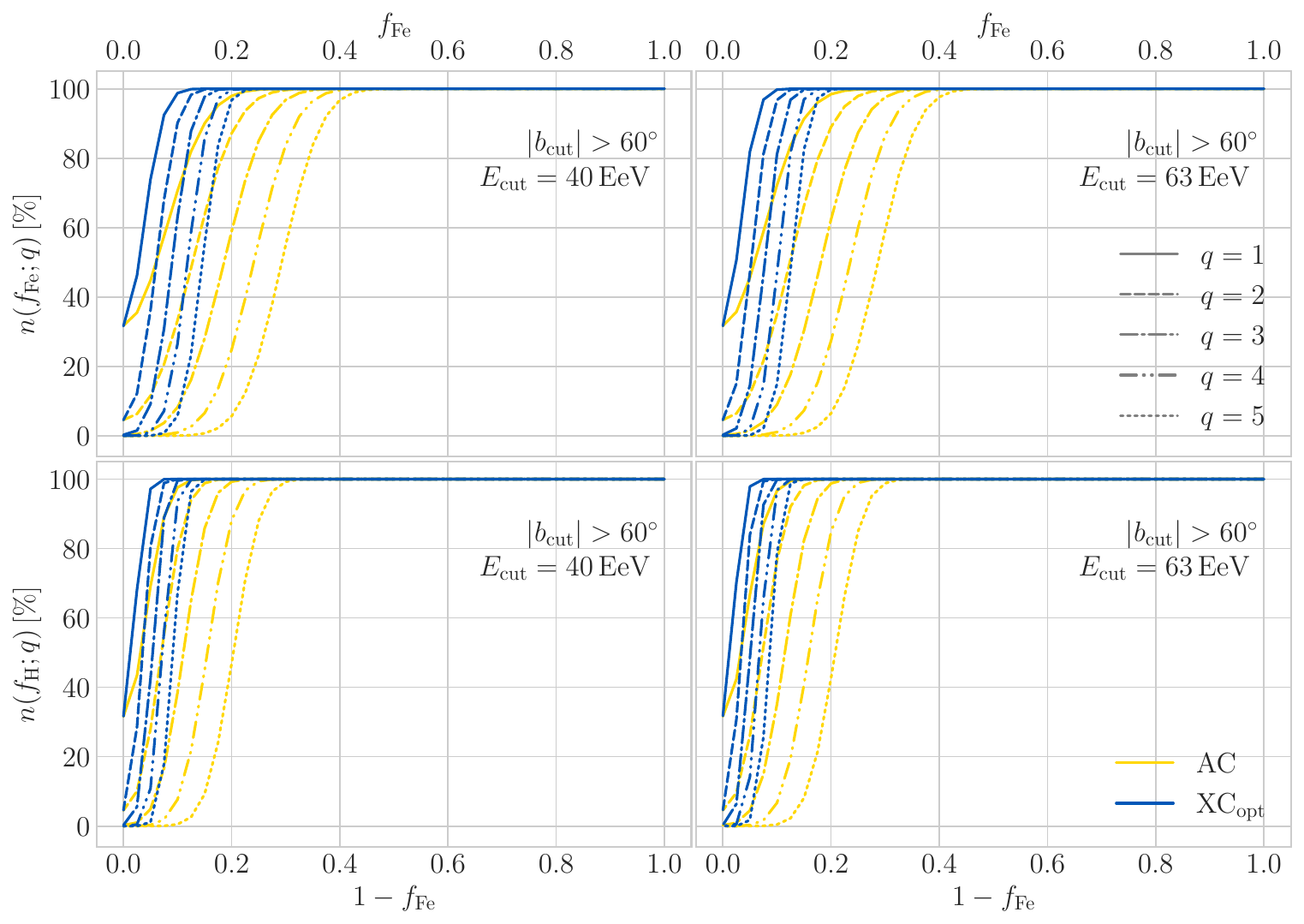}
    \caption{Same as Figure \ref{fig:Ecuts_fFe_Ecuts} but with masking at $|\bcut|=60^\circ$.}
    \label{fig:Ecuts_fFe_Ecuts_mask60}
\end{figure}

Secondly, in order to assess the dependence on choice of sky mask, in \cref{fig:Ecuts_fFe_Ecuts_mask20} and \cref{fig:Ecuts_fFe_Ecuts_mask60} we apply cuts at $|\bcut|=20^\circ$ and $|\bcut|=60^\circ$, respectively. We observe a weak tendency to improve as we cut out more of the sky where the GMF deflections are larger. Hence, once again the loss of UHECR events due to the small sky fraction we observe is more than compensated by the much smaller deflections due to the GMF in that region of the sky. A more realistic test with fully-fledged realisations of the GMF would enable us to maximise this feature to the benefit of the constraining power of our observable.

\begin{figure}
    \centering
    \includegraphics[width=\textwidth]{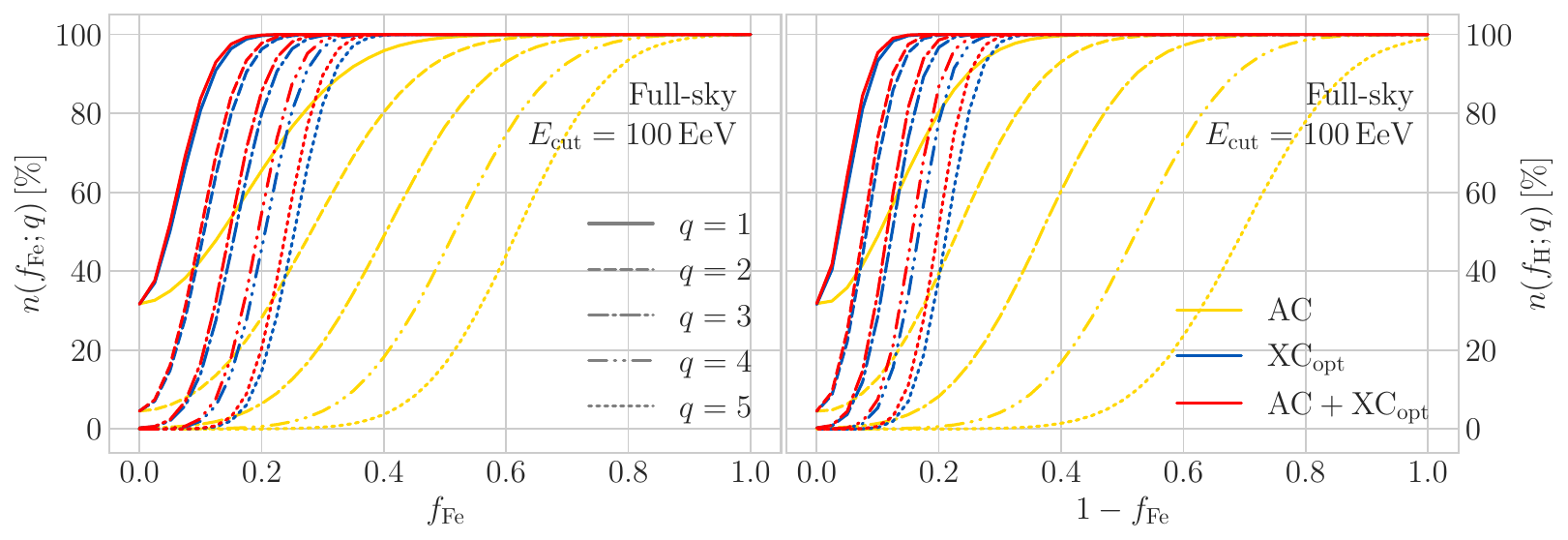}
    \caption{Same as \cref{fig:Ecuts_fFe} but adding, in red, the AC+XC\(_\text{opt}\) according to the TS defined by \cref{eq:chi2cov}.}
    \label{fig:Ecuts_AC_and_XCopt_full}
\end{figure}

Thirdly, we can try to improve our method and results by building the data and theory vectors with all the available measurements, namely the AC and the XC together. Schematically we define \(\bm v_\ell = \{\cS^\text{CR\,CR}_\ell, \cS^\text{CR\,g}_\ell, \cS^\text{g\,g}_\ell\}\), see \citep{Urban:2020szk} for details. In this case the \(\chi^2\) of \cref{eq:chi2} is generalised to
\begin{equation}
\chi^2(F_\mathrm{mix}) = \sum_{\ell\ell'} [ t_\ell(F_\mathrm{mix})-d_\ell]^\intercal\,{\sf\Sigma}_{\ell\ell'}^{-1}\,[ t_\ell'(F_\mathrm{mix})-d_\ell'] \;,
    \label{eq:chi2cov}
\end{equation}
with \({\sf\Sigma}_{\ell\ell'}\) being the covariance matrix. In \cref{fig:Ecuts_AC_and_XCopt_full} we show, in red, the AC+XC\(_\text{opt}\) alongside what we already showed in \cref{fig:Ecuts_fFe}. The improvement brought about by the combination of auto- and cross-correlation is present but not significant.

\begin{figure}
    \centering
    \includegraphics[width=\textwidth]{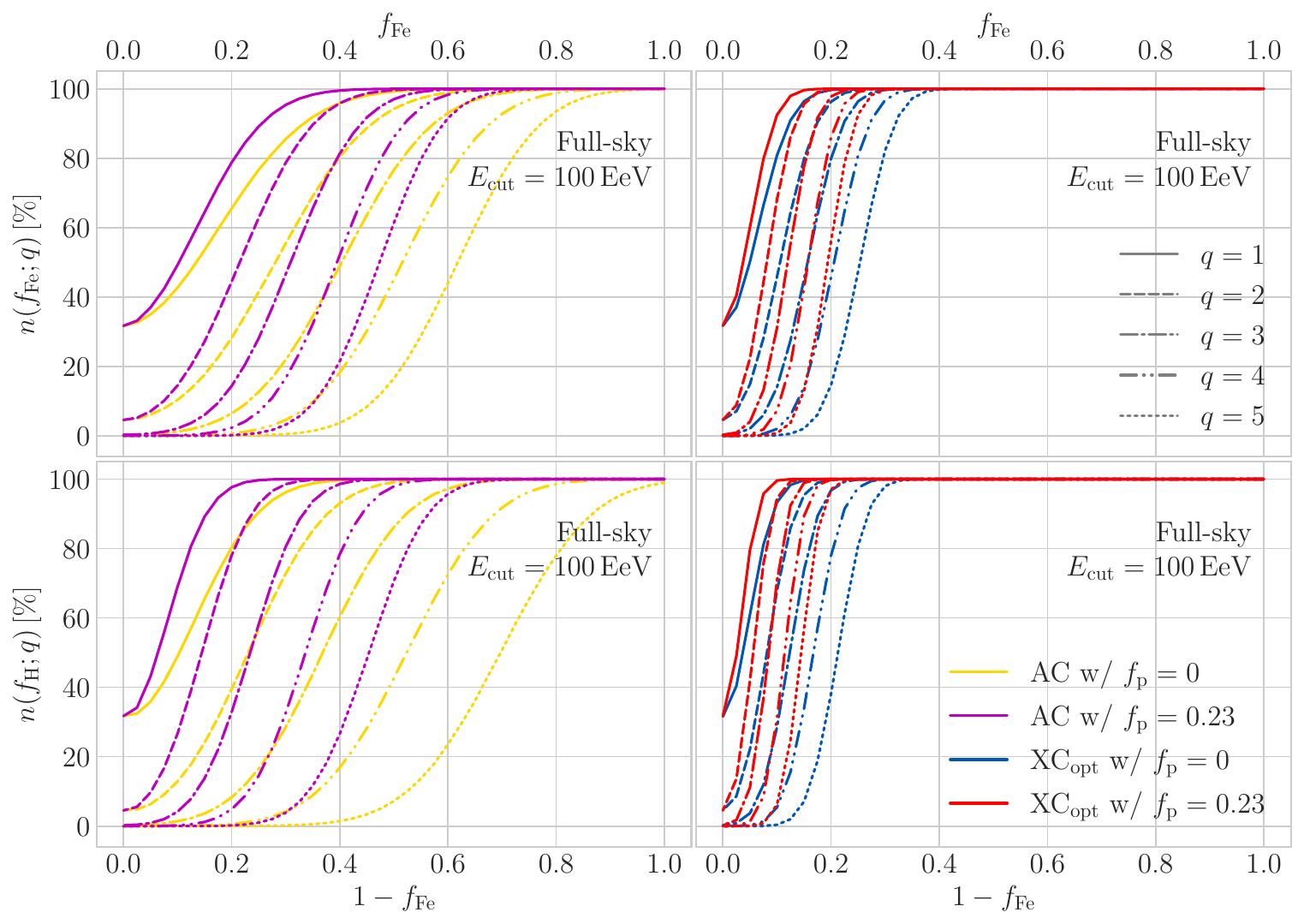}
    \caption{Same as \cref{fig:Ecuts_fFe} but adding the case where \(\Emax=A\Ecut\), which in turn means that \(\fp=0\) (AC: red, XC\(_\text{opt}\): magenta).}
    \label{fig:Ecuts_AC_and_XCopt_Emax}
\end{figure}

Lastly, in order to test the impact of a high-energy injection cutoff on our results, in \cref{fig:Ecuts_AC_and_XCopt_Emax} we compare the AC (left panels) and the XC\(_\text{opt}\) (right panels) for our benchmark model with \(\Ecut=100\,\mathrm{EeV}\) where for \({^{56}}\)Fe we have \(\Emax=\infty\) (AC: yellow, XC\(_\text{opt}\): blue) or \(\Emax=A\Ecut\) (AC: red, XC\(_\text{opt}\): magenta) with \(A=56\). The physical consequence of the high-energy cutoff \(\Emax\) is that in the \(\Emax=A\,\Ecut\) case no child protons with energy \(E\geq\Ecut\) would reach the Earth (instead of making up for approximately \(23\%\) of the flux when there is no high-energy cutoff). Without child protons the \({^{56}}\)Fe spectra become more distinguishable from the \({^1}\)H ones, as evidenced by the smaller fractions of \(\fFe\) (top panels) and \(\fH\) (bottom panels) that can be excluded at a given CL compared to \cref{fig:Ecuts_fFe}. Notice that the total angular power in the case of a childless \({^{56}}\)Fe spectrum on Earth is larger than that with child protons even though the latter, because of the child protons, extends to larger \(\ell\) (smaller angular scales). The reason is that overall the 23\% of child protons contribute to the spectrum less than an equivalent fraction of \({^{56}}\)Fe, so the pure \({^{56}}\)Fe case has more power, and it is easier to tell apart from \({^1}\)H. 

\bibliography{references}

\end{document}